\documentclass{aastex63}
\usepackage{amsmath}

\begin{document}

\title{First ultraviolet outburst detected from ASASSN-18eh strengthens its interpretation as a\\
cataclysmic variable}

\author{Sill Verberne}
\author{David Modiano}
\author{Rudy Wijnands}
\author{on behalf of the TUVO project}
\affiliation{Anton Pannekoek Institute for Astronomy, University of Amsterdam\\
Science Park 904, 1098 XH Amsterdam, Netherlands}

\begin{abstract}
As part of the Transient UV Objects project, we have discovered a new outburst (at the beginning of October 2020) of the candidate cataclysmic variable (CV) ASASSN-18eh using the UV/Optical Telescope aboard the Neil Gehrels Swift Observatory. During the outburst its brightness increased by about 6 mag in UV compared to its brightness in the quiescent state. The properties of this outburst are consistent with it being a dwarf nova, strongly supporting the CV nature of ASASSN-18eh.
\end{abstract}


\section{Introduction} \label{sec:intro}
Cataclysmic variables (CVs) consist of a white dwarf (WD) in close orbit, typically of a few hours \citep[e.g.][]{Knigge_2011}, with usually a red dwarf companion \citep[e.g.][]{Hillman_2020}. The companion overflows its Roche lobe and transfers material to the WD through an accretion disk. The accretion often is not stable but goes through episodic periods (lasting a few to a few tens of days) in which the accretion rate suddenly increases significantly causing bright outbursts in between periods of quiescence (so-called dwarf novae, or DNe, increase in V brightness by about $\sim2-5$ mag). See \citet{Lasota_2001} for a detailed discussion about DNe and the physics involved.

The source ASASSN-18eh has been suggested as a potential CV after an outburst, reaching a V magnitude of 16.1, was observed in February 2018 \citep{Shappee_2014, ASASSN}. As part of our Transient UV Objects (TUVO) project, we have discovered the first outburst in the ultraviolet (UV) of the same source in early October 2020. 

\section{Observations \& Analysis}
The Neil Gehrels Swift Observatory \citep{Gehrels_2004} has been designed to provide rapid multi-wavelength follow-up observations of gamma-ray bursts using, among others, its UV/Optical Telescope \citep[UVOT;][]{Roming_2005}. The observatory offers an excellent opportunity to study transients in the UV given its many repeated pointings of the same field (owing to its high flexibility), and freely accessible daily data supply \citep{Gehrels_2004}.\ 

As part of the TUVO project \citep{Wijnands_2020}, we make use of the \texttt{TUVOpipe} pipeline to search for transients in the data from UVOT \citep{Modiano_2020}. During October 2020, a transient was detected by the pipeline at a position of 14 28 33.52, -46 11 26.5 \citep[J2000; errors of $\lesssim5$ arcsec; see][]{Modiano_2020}, which is consistent with that of ASASSN-18eh, demonstrating that we detected another outburst of this source. This outburst was not reported by either the Zwicky Transient Facility \citep{Bellm_2019} or ASAS-SN \citep{Shappee_2014}.

In total, UVOT has taken 36 exposures of ASASSN-18eh between December 2016 and October 2020. The observations are spread equally between the uw1, um2, and uw2 filters, which have central wavelengths of 2600, 2246, and 1928 \AA\ respectively\footnote{\url{https://www.swift.ac.uk/analysis/uvot/filters.php}}. In nine of the observations (obsIDs 00084507003, 00084507007, and 00084507013; all filters), the source position was right at the boundary of a readout steak caused by a bright star in the field-of-view (FoV) and in another one (obsID 00084507001; only uw1) the source position was located at the edge of the detector and only partially captured. For this reason these exposures have been omitted from the analysis. The resulting total exposure times per filter are $\sim975$, $\sim1284$, and $\sim1284$ seconds for uw1 (8 exposures), um2 (9 exposures), and uw2 (9 exposures) respectively.\

A light curve of the source has been made with the following steps:
firstly, all images were aligned per filter, since UVOT pointings can sometimes be offset by a few arcsec \citep{Poole_2008}. As source extraction region we used a circular region with a radius of 5 arcsec, centered on the source position as determined by our pipeline. For the background extraction region we used a circular region with a radius of $\sim 20$ arcsec in an area close to the target, but without any visible sources. All images were then manually inspected for any potential artifacts such as, for example, the previously mentioned readout streaks close to the source.
Specialized UVOT tools included with the \texttt{HEAsoft} software package\footnote{\url{https://heasarc.gsfc.nasa.gov/docs/software/heasoft/}} (Version 6.28) and the \texttt{HEASARC}'s calibration database (\texttt{CALDB}) system (UVOT version 20190101) were then used to analyse the data. \texttt{Uvotsource}\footnote{\url{https://www.swift.ac.uk/analysis/uvot/mag.php}} was used to determine the magnitudes and fluxes of the source for each exposure.
In individual exposures during quiescence the source was not detected. Therefore, all exposures during quiescence were stacked using \texttt{uvotimsum}\footnote{\url{https://www.swift.ac.uk/analysis/uvot/image.php}} in order to improve the detection limit, allowing us to detect the source in quiescence. The resulting stacked image for each filter was then again passed to \texttt{uvotsource} using the same source and background extraction regions.

\section{Results}
\begin{figure}
    \centering
    \includegraphics[scale=0.95]{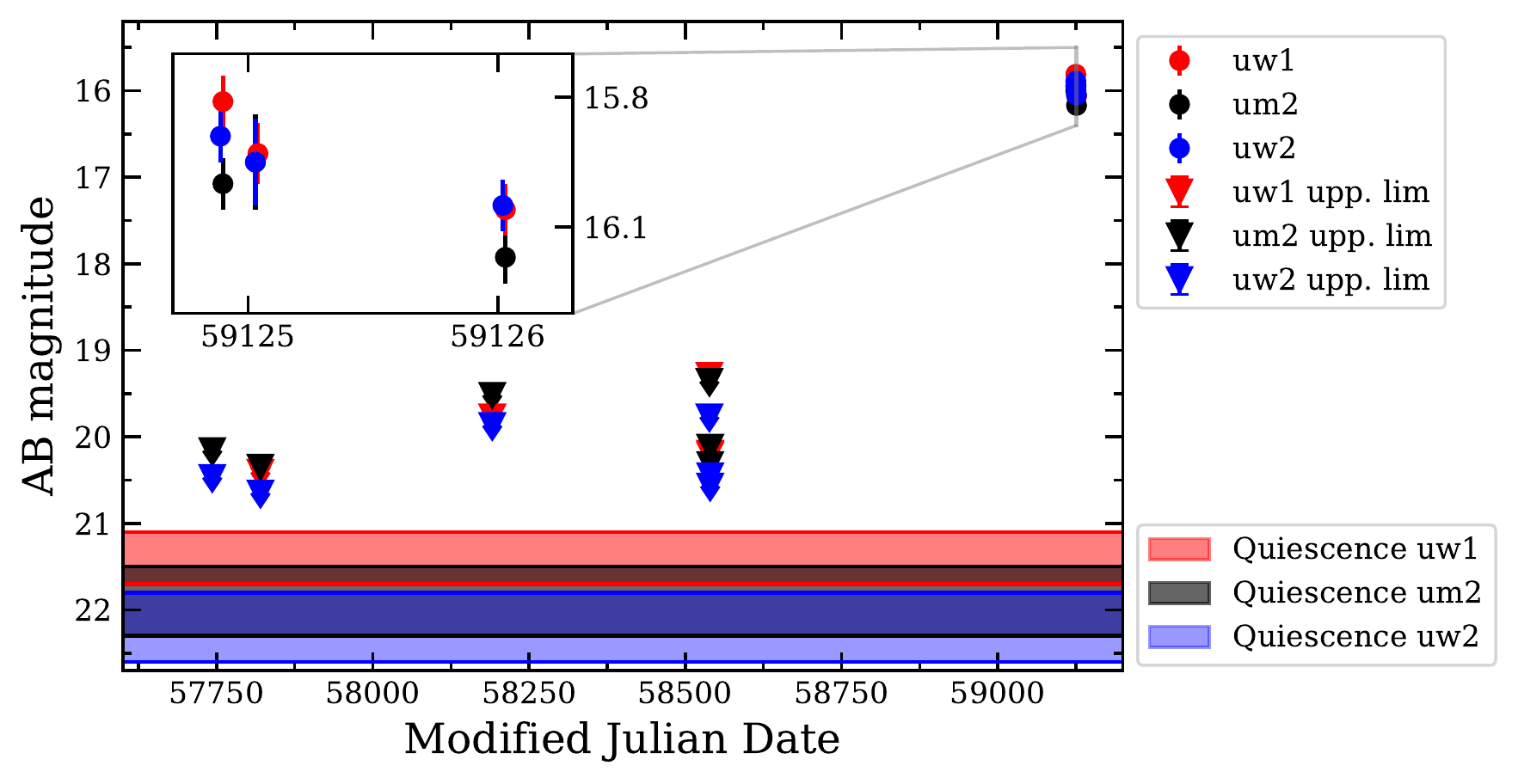}
    \caption{Light curve of ASASSN-18eh showing the observed outburst in the uw1, um2, and uw2 filters. The $1\sigma$ error bars are plotted for the detections in outburst, along with the $1\sigma$ confidence intervals for the detections in quiescence obtained by stacking all the separate observations during which the source was not detected. The time between the last observation in quiescence and the first observation during the outburst is about 1.6 years.}
    \label{fig:lightcurve}
\end{figure}

Figure \ref{fig:lightcurve} shows the light curve of the source in the different filters.
The first exposure that shows the source in outburst was taken at 21:21:36 UTC on 2 October 2020 and the last one at 00:28:48 UTC on 4 October 2020. This gives a lower limit for the outburst duration of just over a day. Unfortunately, no more observations were performed after 4 October 2020, since the field could not be observed anymore due to Solar constraints. This means that the source was still active at the time of our last observation, but further constraints are not possible.\ 

Using the image stacking approach discussed above, we were able to detect the source in quiescence with an AB magnitude and flux density of: \ 

\begin{align*}
    \hspace{-1cm}&\textrm{uw1:} & 21.4 \pm 0.3 \textrm{ mag} && 4.3(\pm1.2)\times10^{-17} \textrm{ erg s}^{-1} \textrm{cm}^{-2} \textrm{\AA}^{-1} && (4.0\sigma \textrm{ detection})\\
    \hspace{-1cm}&\textrm{um2:} & 21.9 \pm 0.4 \textrm{ mag} && 3.7(\pm1.2)\times10^{-17} \textrm{ erg s}^{-1} \textrm{cm}^{-2} \textrm{\AA}^{-1} && (3.1\sigma \textrm{ detection})\\
    \hspace{-1cm}&\textrm{uw2:} & 22.2 \pm 0.4 \textrm{ mag} && 3.4(\pm1.2)\times10^{-17} \textrm{ erg s}^{-1} \textrm{cm}^{-2} \textrm{\AA}^{-1} && (2.9\sigma \textrm{ detection})\\
\end{align*}
This is in contrast with the brightest AB magnitudes and flux densities measured during outburst (i.e. observed during the first outburst observations) of:\ 
\begin{align*}
    \hspace{-1cm}&\textrm{uw1:}& 15.81\pm0.06 \textrm{ mag}&& 8.32(\pm0.46)\times10^{-15}\textrm{ erg s}^{-1} \textrm{cm}^{-2} \textrm{\AA}^{-1}\\
    \hspace{-1cm}&\textrm{um2:}& 15.95\pm0.11 \textrm{ mag}&& 9.24(\pm0.72)\times10^{-15}\textrm{ erg s}^{-1} \textrm{cm}^{-2} \textrm{\AA}^{-1}\\
    \hspace{-1cm}&\textrm{uw2:}& 15.89\pm0.06 \textrm{ mag}&& 1.21(\pm0.06)\times10^{-14}\textrm{ erg s}^{-1} \textrm{cm}^{-2} \textrm{\AA}^{-1}\\ 
\end{align*}

Therefore, the outburst had an amplitude of at least $\sim6$ mag across all filters. It is also noteworthy that all subsequent measurements during the outburst show a decreasing brightness, which indicates that the peak brightnesses were likely higher than the above quoted ones.

\section{Conclusions}
We have presented observations of the first detected UV outburst of the candidate CV ASASSN-18eh which was found during October 2020. The UV intensity during the outburst increased by at least $\sim6$ mag with a minimum duration of just over a day. Unfortunately, more observations during the October 2020 outburst could not be obtained because of Solar constraints. The next time observations can be obtained is in late December 2020 by which time it is expected that the source has decayed into quiescence again.
The maximum UV brightnesses we measured during the October 2020 outburst are slightly higher compared to the brightness in V measured by ASAS-SN during the outburst at the end of February 2018. This is consistent with DNe emitting a large fraction (if not most) of their energy in the UV \citep[e.g.][]{Giovannelli_2008, parikh_2019}, although we note that we might have missed the peak during the observations making stringent inferences difficult.
The previous known outburst (and so far the only one) of the source was in February 2018, meaning that the recurrence time is not more than 2.7 years, although very likely it is shorter because additional outbursts were likely missed by surveying transient facilities. Typical recurrence times of DNe range from days to decades \citep{Belloni_2016}, depending mostly on the mass ratio between the WD and its companion \citep{Patterson_2011}. 
These characteristics (i.e. outburst amplitude, duration, and recurrence time) are consistent with the outburst being a DN outburst of a CV, which would confirm the CV nature of the source.

%

\vspace{5mm}
\facilities{Swift (UVOT)}





\bibliographystyle{aasjournal}
\bibliography{references}



\end{document}